\documentclass[prl,twocolumn]{revtex4}

\usepackage{epsfig}

\renewcommand{\vec}[1]{{\rm\bf #1}}

\newcommand{\ep}{\epsilon}

\newcommand{\unitmatrix}{\openone}

\begin{document}

\title{Effect of inelastic collisions
on multiphonon Raman scattering in graphene}
\author{D.~M.~Basko}\email{basko@phys.columbia.edu}
\affiliation{Physics Department, Columbia University, New York, NY
  10027, USA}
%\author{I.~L.~Aleiner}
%\affiliation{Physics Department, Columbia University, New York, NY
%10027, USA}

\begin{abstract}
We calculate the probabilities of two- and four-phonon Raman
scattering in graphene and show how the relative intensities of the
overtone peaks encode information about relative rates of different
inelastic processes electrons are subject to.
If the most important processes are
electron-phonon and electron-electron scattering, the rate of the
latter can be deduced from the Raman spectra.
\end{abstract}

\maketitle

{\em Introduction.}--- Collisions of Dirac electrons are
qualitatively different from those of electrons in conventional
metals: energy and momentum conservation leave no phase space for
relaxation of a quasiparticle excited above the Dirac vacuum.
Electron collisions in graphene and related compounds continue to
be a subject of theoretical
studies~\cite{Guinea96,Rubio,DasSarma}. Thus, any experimental
information on collisions of Dirac electrons would be extremely
valuable. However, to experimentally separate contributions from
different mechanisms to electron lifetime is a hard task. The
present work suggests a way to separate electron-phonon and
electron-electron contributions to the electron lifetime by
analyzing Raman spectra.

Raman spectrum of graphene consists of distinct peaks
corresponding to different optical phonon branches as well as
their overtones. Thus, Raman scattering measurements represent a
powerful experimental tool for studying phonon modes, as well as
their interaction with electrons (since
%graphene is a non-polar crystal, phonons do not interact with light
%directly, so
electronic excitations are involved in Raman scattering as
intermediate states). Indeed, electron-phonon interaction and
Raman scattering in graphene has attracted a great deal of
interest, both experimental~\cite{Ferrari,Gupta,Graf,Pisana,Jun}
and theoretical~\cite{Lazzeri,Guinea}. Here we show how even more
information can be extracted from Raman spectra.

\begin{figure}
\includegraphics[width=0.45\textwidth]{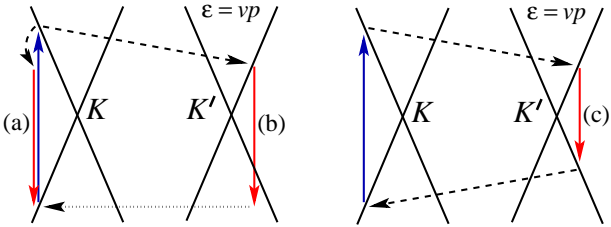}
\caption{\label{fig:res} Schematic representation of the role of
  electron dispersion (Dirac cones, shown by solid lines) in Raman
  scattering: (a)~one-phonon $G$~peak, (b)~impurity-assisted
  one-phonon $D$~peak, (c)~two-phonon $D^*$~peak. Vertical solid
  arrows represent interband electronic transitions accompanied by
  photon absorption or emission (photon wave vector is neglected),
  dashed arrows represent phonon emission, the horizontal dotted arrow
  represents impurity scattering.}
\end{figure}

{\em Qualitative picture.}--- Photon wave vector is
negligible, so momentum conservation requires that Raman
scattering on one intervalley phonon must be impurity-assisted
[process~(b) in Fig.~\ref{fig:res}, giving rise to the so-called
$D$~Raman peak]. $D$~peak is absent in the experimental Raman
spectrum of graphene~\cite{Ferrari}, so impurity
scattering is negligible in these samples, and is disregarded hereafter.

Looking at the intermediate electronic states
%involved in the Raman scattering
(Fig.~\ref{fig:res}), we notice that for one-phonon scattering
[processes (a),~(b)] at least one intermediate state must be
virtual, since energy and momentum conservation cannot be
satisfied simultaneously. For the two-phonon scattering
[process~(c)] all electronic states can be real. We emphasize the
qualitative difference between the {\em fully resonant}
process~(c) and the double-resonant~\cite{ThomsenReich}
process~(b), where one intermediate state is still virtual.

Obviously, our argument can be extended to all multi-phonon
processes with odd and even number of phonons involved: in order
to annihilate radiatively, the electron and the hole must have
opposite momenta; if the total number of emitted phonons is odd,
the electron and the hole must emit a different number of phonons,
which is incompatible with energy conservation in all processes.
%This argument implies that electron-hole asymmetry and
%optical phonon dispersion are much smaller than the phonon
%frequency, which is true.

Real even-phonon processes can be viewed in the following way. The
incident photon creates an electron and a hole -- real
quasiparticles which can participate in various scattering
processes. If the electron emits a phonon with a
momentum~$\vec{q}$, the hole emits a phonon with the momentum
$-\vec{q}$, and after that the electron and the hole recombine
radiatively, the resulting photon will contribute to the
two-phonon Raman peak. If they do not recombine at this stage, but
each of them emits one more phonon, and they recombine afterwards,
the resulting photon will contribute to the four-phonon peak, {\it
etc.}

Besides phonon emission and radiative recombination, electron and hole
are subject to other inelastic scattering processes, which can also
be viewed as emission of some excitations of the system.
%In principle,
%Raman spectrum should also contain the contrubution from these
%excitations, which are left in the system after the radiative
%recombination of the electron and the hole.
The key point is that for
real quasiparticles, the probability to undergo a scattering
process~$\alpha$ is determined by the ratio of corresponding
scattering rate $2\gamma_\alpha$ to the total
scattering rate $2\gamma\equiv\sum_\alpha{2}\gamma_\alpha$, not by
history. This probability determines the relative {\em
  frequency-integrated} intensity of the corresponding feature in the
Raman spectrum. Thus, the ratio of integrated intensity $I_{2n+2}$ of
the Raman peak corresponding to $2n+2$ phonons to that for
$2n$~phonons ($I_{2n}$) must be proportional to
%\begin{equation}
%\frac{I_{2n+2}}{I_{2n}}\propto\left(\frac{\gamma_{ph}}\gamma\right)^2.
%\end{equation}
$(\gamma_{ph}/\gamma)^2$, where
$2\gamma_{ph}$ is the rate of emission of each of the two
phonons, and the square comes from the phonon emission by the
electron and the hole.

In the Raman spectrum of graphene two two-phonon peaks are seen:
the so-called $D^*$ peak near $2\omega_{A_1}=2650\:\mathrm{cm}^{-1}$,
and the $G^*$ peak near $2\omega_{E_2}=3250\:\mathrm{cm}^{-1}$,
corresponding to scalar $A_1$~phonons from the
vicinity of the $K$~point of the first Brillouin zone, and to
pseudovector $E_2$~phonons from the vicinity of the $\Gamma$~point,
respectively. The $D^*$ peak is more intense, thus the most
interesting for practical purposes is to compare the intensities of
$D^*$~and its four-phonon overtone at 5300~cm$^{-1}$, which we will
call~$2D^*$:
\begin{equation}\label{answer=}
I_{2D^*}/I_{D^*}=0.14\left(\gamma_{A_1}/\gamma\right)^2.
\end{equation}
The coefficient~0.14 was obtained by direct calculation assuming
$\omega_{in}\gg\omega_{A_1}\gg\gamma$, where
$\omega_{in}$~is the incident photon frequency. $2\gamma_{A_1}$ is
the rate of emission of the $A_1$~phonon. It depends on the electron
energy, which can be taken to be $\omega_{in}/2$.
Eq.~(\ref{answer=}) represents the main result of this paper. Let
us now discuss its practical implication.

In graphene, the most obvious competitor of the phonon emission is
the electron-electron scattering: the optically excited electron
can kick out another one from the Fermi sea, i.~e., to emit
another electron-hole pair (for an electron above the Dirac vacuum
the phase space for an intravalley collision is zero, so the
collision has to be intervalley or
impurity-assisted~\cite{Guinea96}). Thus, Raman spectrum should
contain contribution from electron-hole pairs; however, their
spectrum extends all the way to the energy of the photo-excited
electron (optical energy) in a completely featureless way. Thus,
it cannot be distinguished from the parasitic background which is
always subtracted in the analysis of Raman spectra, and cannot be
seen in the Raman spectrum directly. However, assuming
$\gamma=\gamma_{A_1}+\gamma_{E_2}+\gamma_{ee}$, where
$2\gamma_{E_2}$ is the rate of the $E_2$~phonon emission, and
$2\gamma_{ee}$ is the electron-electron collision rate, one can
extract the value of $\gamma_{ee}$ from the experimental data
using Eq.~(\ref{answer=}), relative to phonon emission rates. More
precisely, in this way one obtains the rate of all inelastic
scattering processes where the electron loses energy far exceeding
the phonon energy.

Note that arguments leading to
$I_{2n+2}/I_{2n}\propto(\gamma_{ph}/\gamma)^2$ are not specific for
graphene; in fact, this is nothing but Breit-Wigner formula, applied
once for the electron and once for the hole.
Multi-phonon
Raman scattering has been studied in wide-gap semiconductors both
experimentally~\cite{Damen,Porto} (up to ten phonons were seen in
the Raman spectra of CdS), and theoretically~\cite{Varma,Zeyher}.
In a wide-gap semiconductor an optically excited electron does not
have a sufficient energy to excite another electron across the
gap, so the electron-electron channel is absent. In addition,
interaction with only one phonon mode is dominant, so the ratios
of subsequent peaks are represented by a sequence of fixed
numbers. The simple band structure
%(one valley for CdS in contrast to two valleys for graphene)
allowed a calculation of the whole
sequence. %To do the same for graphene is problematic,
%because of the more complicated band structure,
%A more complicated electronic band structure in graphene
%makes it problematic to calculate the whole sequence,
For graphene, we restrict ourselves to the calculation leading to
Eq.~(\ref{answer=}).

\begin{table}
\begin{tabular}[t]{|c|c|c|c|c|c|c|} \hline
$C_{6v}$ & $E$ & $C_2$ & $2C_3$ & $2C_6$ & $\sigma_{a,b,c}$ &
$\sigma_{a,b,c}'$
\\ \hline\hline $A_1$ & 1 & 1 & 1 & 1 & 1 & 1 \\ \hline $A_2$ & 1
& 1 & 1 & 1 & $-1$ & $-1$ \\ \hline $B_2$ & 1 & $-1$ & 1 & $-1$ &
$1$ & $-1$ \\ \hline $B_1$ & 1 & $-1$ & 1 & $-1$ & $-1$ & $1$ \\
\hline $E_1$ & 2 & $-2$ & $-1$ & $1$ & 0 & 0 \\ \hline $E_2$ & 2 &
2 & $-1$ & $-1$ & 0 & 0 \\ \hline\end{tabular}\hspace{1cm}
%begin{tabular}[t]{|c|c|c|c|}
%\hline $C_{3v}$ & $E$ & $2C_3$ & $\sigma_{a,b,c}'$ \\ \hline\hline
%$A_1$ & 1 & 1 & 1 \\ \hline $A_2$ & 1 & 1 & $-1$ \\ \hline $E$ & 2
%& $-1$ & 0 \\ \hline
%\end{tabular}
\caption{Irreducible representations of the group $C_{6v}$ and
their characters.\label{tab:C6vC3v}}
\end{table}

\begin{table}
\begin{tabular}{|c|c|c|c|c|c|c|} \hline
irrep & %$A_1$ & $B_1$ & $A_2$ & $B_2$ & $E_1$ & $E_2$ &
$A_1$ & $B_1$ & $A_2$ & $B_2$ & $E_1$ & $E_2$ \\ \hline
\multicolumn{7}{|c|}{valley-diagonal matrices}\\
\hline
matrix & $\unitmatrix$ & $\Lambda_z$ & $\Sigma_z$ &
$\Lambda_z\Sigma_z$ & $\Sigma_x,\,\Sigma_y$ &
$-\Lambda_z\Sigma_y,\Lambda_z\Sigma_x$ \\ \hline
\multicolumn{7}{|c|}{valley-off-diagonal matrices} \\
\hline
matrix & $\Lambda_x\Sigma_z$ &
$\Lambda_y\Sigma_z$ & $\Lambda_x$ & $\Lambda_y$ &
$\Lambda_x\Sigma_y,-\Lambda_x\Sigma_x$ &
$\Lambda_y\Sigma_x,\Lambda_y\Sigma_y$ \\ \hline
\end{tabular}
\caption{Classification of $4\times{4}$ hermitian matrices
according to irreducible representations of the $C_{6v}$~group.
\label{tab:matrices}}
\end{table}

{\em Model.}--- We measure the single-electron energies from the
Fermi level of the undoped (half-filled) graphene. The Fermi
surface of undoped graphene consists of two points, called
$K$~and~$K'$. Graphene unit cell contains two atoms, each of them
has one $\pi$-orbital, so there are two electronic states for each
point of the first Brillouin zone (we disregard the electron
spin). Thus, there are exactly four electronic states with zero
energy. An arbitrary linear combination of them is represented by
a 4-component column vector~$\psi$. States with low energy are
obtained by including a smooth position dependence
$\psi(\vec{r})$, $\vec{r}\equiv(x,y)$. The low-energy hamiltonian
has the Dirac form:
\begin{equation}\label{Hel=}
\hat{H}_{el}=\int{d}^2\vec{r}\,\hat\psi^\dagger(\vec{r})\,
(-iv\vec\Sigma\cdot\vec\nabla)\,\hat\psi(\vec{r}).
\end{equation}
We prefer not to give the explicit form of the isospin matrices
$\vec\Sigma\equiv(\Sigma_x,\Sigma_y)$, which depends on the choice
of the basis (specific arrangement of the components in the
column~$\psi$). We only note that all 16 generators of the $SU(4)$
group, forming the basis in the space of $4\times{4}$ hermitian
matrices, can be classified according to the irreducible
representations of~$C_{6v}$, the point group of the
graphene crystal (Tables \ref{tab:C6vC3v} and~\ref{tab:matrices}).
They can be represented as products of two mutually commuting algebras
of Pauli matrices $\Sigma_x,\Sigma_y,\Sigma_z$ and
$\Lambda_x,\Lambda_y,\Lambda_z$~\cite{Falko,AleinerEfetov}, which
fixes their algebraic relations. By
definition, $\Sigma_x,\Sigma_y$ are the matrices, diagonal in the
$K,K'$ subspace, and transforming according to the
$E_1$~representation of~$C_{6v}$.

\begin{figure}
\includegraphics[width=7cm]{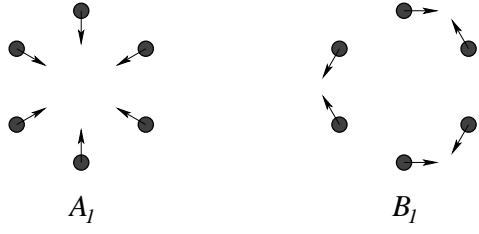}
\caption{\label{fig:phonons} Phonon modes responsible for the
$D^*$
  Raman peak.}
\end{figure}

We restrict our attention to scalar phonons with wave vectors
close to $K$ and $K'$ points -- those responsible for the $D^*$
Raman peak. The two real linear combinations of the modes at $K$
and $K'$ points transform according to $A_1$ and $B_1$
representations of~$C_{6v}$ and are shown in
Fig.~\ref{fig:phonons}. We take the magnitude of the carbon atom
displacement as the normal coordinate for each mode, denoted by
$u_a$ and~$u_b$, respectively. Upon quantization of the phonon
field, $\hat{u}_a,\hat{u}_b$ and the phonon hamiltonian
$\hat{H}_{\mathrm{ph}}$ are expressed in terms of the creation and
annihilation operators
$\hat{b}^\dagger_{\vec{q}\mu},\hat{b}_{\vec{q}\mu}$, $\mu=a,b$, as
\begin{equation}
\hat{u}_\mu(\vec{r})=
%L_xL_y\int\frac{d^2\vec{q}}{(2\pi)^2}
\sum_{\vec{q}}
\frac{\hat{b}_{\vec{q}\mu}e^{i\vec{q}\vec{r}}
%+\hat{b}_{\vec{q}\mu}^\dagger e^{-i\vec{q}\vec{r}}
+\mathrm{h.c.}
}
{\sqrt{2NM\omega_{A_1}}},\quad
\hat{H}_{\mathrm{ph}}=\sum_{\vec{q},\mu}
%L_xL_y\int\frac{d^2\vec{q}}{(2\pi)^2}
%\sum_{\mu=a,b}
\omega_{A_1}\hat{b}_{\vec{q}\mu}^\dagger\hat{b}_{\vec{q}\mu}.
%&&\sum_{\vec{q}}\equiv L_xL_y\int\frac{d^2\vec{q}}{(2\pi)^2}.\nonumber
\end{equation}
The crystal is assumed to have the area $L_xL_y$, and to contain
$N$~carbon atoms of mass~$M$. The $\vec{q}$~summation is performed as
$\sum_\vec{q}\to{L}_xL_y\int{d}^2\vec{q}/(2\pi)^2$. ``h.c.''
stands for hermitian conjugate.
By symmetry, in the electron-phonon interaction hamiltonian~\cite{Heph}
the normal displacements $u_\mu$ couple to the corresponding
valley-off-diagonal $4\times{4}$ matrices from
Table~\ref{tab:matrices}:
\begin{equation}\label{Heph=}
\hat{H}_{int}=3F_{A_1}\int{d}^2\vec{r}\,\hat\psi^\dagger(\vec{r})
\left[\hat{u}_a(\vec{r})\Lambda_x\Sigma_z
+\hat{u}_b(\vec{r})\Lambda_y\Sigma_z\right] \hat\psi(\vec{r}).
\end{equation}

Interaction with light is obtained from the Dirac
hamiltonian~(\ref{Hel=}) by replacement
$\vec\nabla\to\vec\nabla-i(e/c)\hat{\vec{A}}$, where the vector
potential~$\hat{\vec{A}}$ is expressed in terms of creation and
annihilation operators
$\hat{a}^\dagger_{\vec{Q},\ell},\hat{a}_{\vec{Q},\ell}$ of
three-dimensional photons in the quantization volume $V=L_xL_yL_z$,
labeled by the wave vector~$\vec{Q}$ and two transverse
polarizations $\ell=1,2$ with unit vectors $\vec{e}_{\vec{Q},\ell}$:
\begin{equation}
\hat{\vec{A}}(\vec{r})=
\sum_{\vec{Q},\ell}\sqrt{\frac{2\pi{c}}{VQ}}
\left(\vec{e}_{\vec{Q},\ell}\hat{a}_{\vec{Q},\ell}e^{i\vec{Q}\vec{r}}
%+\vec{e}^*_{\vec{Q},\ell}\hat{a}^\dagger_{\vec{Q},\ell}
%e^{-i\vec{Q}\vec{r}}
+\mathrm{h.c.}
\right).
\end{equation}

\begin{figure}
\includegraphics[width=8cm]{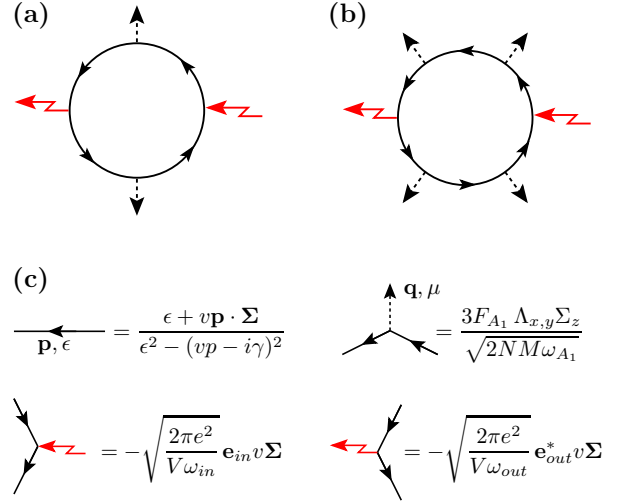}
\caption{\label{fig:diagrams} (a), (b) Fully resonant diagrams giving
  the matrix   element $\mathcal{M}$ for the two- and four-phonon
  Raman scattering, respectively. Other possible diagrams (not shown)
  are not fully resonant, and are smaller by a factor
  $\sim\gamma/\omega_{A_1}$.
 (c)~Analytical expressions, corresponding to lines and
  vertices of the diagrams. In the electron loop integration over the
  internal momentum and energy, as well as tracing over $4\times{4}$
  matrix structure should be performed.
}
\end{figure}

{\em Calculation.}---
$\mathcal{M}(\vec{q}_1,\mu_1;\ldots;\vec{q}_n,\mu_n)$, the matrix
element of the transition from the state with one incoming photon
(frequency~$\omega_{in}$, polarization~$\vec{e}_{in}$) into the
state with one outgoing photon (frequency~$\omega_{out}$,
polarization~$\vec{e}_{out}$) and $n$~phonons (modes
$\mu_1,\ldots,\mu_n$, wave vectors $\vec{q}_1,\ldots,\vec{q}_n$),
is calculated as shown in Fig.~\ref{fig:diagrams}. We emphasize
the necessity to include the inelastic broadening~$\gamma$ in the
electronic Green's functions, as the dominant contribution to the
integral comes from regions where the denominators are small
($\sim\gamma$, corresponding to real electrons and holes, as
discussed above). Given the matrix element, we sum over final
photon and phonon states, and express the absolute dimensionless
probability of $n$-phonon Raman scattering as
\begin{eqnarray}\nonumber
I_n&=&\frac{V^2}{c^2}\,\frac{L_xL_y\omega_{out}^2}{2\pi^2{c}^2}
\frac{1}{n!}\times\\ &&\times\sum_{\vec{q}_1+\ldots+\vec{q}_n=0}
\sum_{\{\mu_i\}} \left|\sum_{\mathcal{P}}\mathcal{M}\!
\left({\mathcal{P}\{\vec{q}_i,\mu_i\}}\right)\right|^2,
\label{Inph=}
\end{eqnarray}
where $\mathcal{P}$ denotes permutations of phonon
arguments.

For the $D^*$~peak %(2700~cm$^{-1}$)
the diagram in Fig.~\ref{fig:diagrams}a gives
\begin{equation}\label{M2ph=}
\mathcal{M}(\vec{q})=
\frac{\pi{e}^2/V}{\sqrt{\omega_{in}+\omega_{out}}}\,
\frac{9F_{A_1}^2/4}{NM\omega_{A_1}}\, %\times\nonumber\\&&\times
\frac{[\vec{e}_{\vec{q}}\times\vec{e}_{in}]_z
[\vec{e}_{\vec{q}}\times\vec{e}_{out}^*]_z}
{[v(q-q_{bs})-2i\gamma]^{3/2}},
\end{equation}
where $\vec{e}_{\vec{q}}\equiv\vec{q}/|\vec{q}|$. The value
$q_{bs}=(\omega_{in}+\omega_{out})/(2v)$, around which
expression~(\ref{M2ph=}) is strongly peaked, corresponds to
backscattering of the electron and the hole by the phonons. This
sharply peaked dependence cannot be obtained from pure symmetry
considerations, which just prescribe vanishing of the matrix
element when the scattering angle $\varphi=0$~\cite{Maultzsch}.
Its origin is the quasiclassical nature of the electron and hole
motion~\cite{thickpaper}, the dispersion of~$\varphi$ being
determined by quantum diffraction:
$|\varphi-\pi|\sim\sqrt{\gamma/\omega_{in}}\ll{1}$. One
consequence of this peaked dependence is that the width of the
$D^*$~peak is determined by the electron and phonon lifetimes
only, not by the phonon dispersion. Besides, it should lead to a
significant polarization anisotropy of the $D^*$
peak~\cite{thickpaper}. Here we simply sum over the polarizations;
substituting Eq.~(\ref{M2ph=}) into Eq.~(\ref{Inph=}) and
approximating $\omega_{out}\approx\omega_{in}$, we obtain:
\begin{equation}\label{ID*=}
I_{D^*}=
\frac{(e^2/c)^2}{48\pi}\frac{v^2}{c^2}\frac{\omega_{in}^2}{\gamma^2}
\left[\frac{9F_{A_1}^2}{M\omega_{A_1}v^2}
\frac{\sqrt{27}a^2}4\right]^2.
\end{equation}
We can allow for trigonal warping and electron-hole asymmetry
terms in the dispersion of electrons and holes:
$\pm{v}p\to\pm(vp+\alpha_3p^2\cos{3}\varphi_{\vec{p}})+\alpha_0p^2$,
where $\varphi_{\vec{p}}$ is the polar angle of~$\vec{p}$,
tight-binding model gives $\alpha_3=va/4$
($v\approx{10}^8\:\mathrm{cm/s}$, $a\approx{1}.42\:\mbox{\AA}$),
and
$\alpha_0(1\:\mbox{eV})^2/v^2\sim{0}.1\:\mbox{eV}$~\cite{DresselhausBook}.
The relative corrections to expression~(\ref{ID*=}) from these
terms are $(9/8)(\alpha_3\omega_{A_1}/v^2)^2\sim{10}^{-4}$ and
$-[\alpha_0(\omega_{in}\omega_{A_1}/v^2)/(2\gamma)]^2/2%
\sim{10}^{-4}(\omega_{in}/2\gamma)^2$.

Evaluation of the diagram in Fig.~\ref{fig:diagrams}b gives the
intensity of the four-phonon Raman peak:
\begin{equation}\label{I2D*=}
I_{2D^*}=0.088\,
\frac{(e^2/c)^2}{64\pi^4}\frac{v^2}{c^2}\frac{\omega_{in}^4}{\gamma^4}
\left[\frac{9F_{A_1}^2}{M\omega_{A_1}v^2}
\frac{\sqrt{27}a^2}4\right]^4.
\end{equation}

\begin{figure}
\includegraphics[width=4cm]{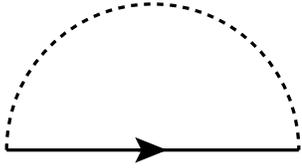}
\caption{\label{fig:selfenrg} The self-energy diagram describing the
  phonon emission. The dashed line represents the phonon Green's
  function $D(\vec{q},\omega)=2\omega_{A_1}/[\omega^2-(\omega_{A_1}-io)^2]$.
}
\end{figure}

Finally, we express the Raman scattering probabilities in terms of
the phonon emission rate. The latter is given by the imaginary
part of the self-energy %~\cite{Guinea1981}
(Fig.~\ref{fig:selfenrg}):
\begin{equation}\label{gammaph=}
\gamma_{A_1}(\ep)= \frac{9F^2_{A_1}}{M\omega_{A_1}v^2}
\frac{\sqrt{27}a^2}4\,
\frac{|\ep|-\omega_{A_1}}{4}\,\theta(|\ep|-\omega_{A_1}),
\end{equation}
where $\theta(\ep)$ is a step
function. Eqs.~(\ref{ID*=}),~(\ref{I2D*=}) and Eq.~(\ref{gammaph=})
with $\ep=\omega_{in}/2$ give Eq.~(\ref{answer=}).

Instead of conclusion, we quote an experimental value for the
ratio $I_{D^*}/I_{2D^*}\approx{40}$~\cite{expdata}, so
Eq.~(\ref{answer=}) gives $\gamma_{A_1}/\gamma\approx{0}.42$. We
assume $\gamma=\gamma_{A_1}+\gamma_{E_2}+\gamma_{ee}$ and note
that the emission rate $2\gamma_{E_2}$ of $E_2$~phonons is
described by Eq.~(\ref{gammaph=}), with the replacements
$F_{A_1}\to{F}_{E_2}$ (the corresponding coupling constant),
$\omega_{A_1}\to\omega_{E_2}$. In the tight-binding model
$F_{A_1}=F_{E_2}$ (for the normalization of the phonon
displacements chosen here), which agrees with a DFT
calculation~\cite{Piscanec} up to~1\%. The assumption
$F_{A_1}=F_{E_2}$ gives $\gamma_{ee}\approx{0}.22\,\gamma$.

On the other hand, $F_{A_1}$~and~$F_{E_2}$ are not related by any
symmetry. For intensities of the two-phonon Raman peaks
$D^*$~and~$G^*$ our calculation gives
$I_{D^*}/I_{G^*}=2(F_{A_1}/F_{E_2})^4(\omega_{E_2}/\omega_{A_1})^2$,
the experimental value being
$I_{D^*}/I_{G^*}\approx{20}$~\cite{Ferrari,expdata}. This suggests
$F_{A_1}/F_{E_2}\approx{1}.6$, in significant disagreement with
the tight-binding model prescription. Substituted in
Eq.~(\ref{answer=}), it gives $\gamma_{ee}\approx{0}.44\,\gamma$,
which agrees with the calculated
$2\gamma_{ee}\approx{10}$~meV~\cite{Rubio} and the total $2\gamma$
measured by time-resolved photoemission spectroscopy (20~meV in
Ref.~\cite{Gao}, $25$~meV in Ref.~\cite{Moos}, all values taken
for $\ep=\omega_{in}/2=1$~eV). Even though a recent ARPES
measurement gives a significantly larger value for
$2\gamma\sim{100}\:\mbox{meV}$~\cite{Rotenberg}, the issue of
validity of the tight-binding model for electron-phonon coupling
seems to deserve further investigation.

The author thanks I.~L.~Aleiner and J.~Yan for stimulating
discussions and critical reading of the manuscript, and Y.~Wu for
sharing unpublished experimental results.


\begin{thebibliography}{99}
\bibitem{Heph}
Hamiltonian~(\ref{Heph=}) can be derived microscopically, e.~g., in
the tight-binding model. Then $F\equiv\partial{t}_0/\partial{a}$,
where $t_0$~is the nearest-neighbor coupling matrix element, and
$a$~is the bond length. Our calculation, however, does not rely on the
tight-binding approximation.
\end{thebibliography}
\end{document}